# Propagation of an ultrashort electromagnetic pulse in solid-state plasma

V.A. Astapenko, E.S. Manuilovich

*Moscow Institute of Physics and Technology*

The development of technology for generation of ultrashort pulses (USP) of a specified shape makes urgent the problem of their interaction with a medium [1-3].

This work is dedicated to the study of change of the shape of an USP during its propagation in solid-state plasma.

As an example, we will consider the propagation in bulk silver of a cosine wavelet pulse of the form [4]:

$$E_c(t) = \frac{2}{\sqrt{3}\,\pi^{1/4}} E_0 \left(1 - \frac{t^2}{\Delta t^2}\right) \exp\left(-\frac{t^2}{2\Delta t^2}\right) \qquad (1)$$

that has a proper name "Mexican hat" (Fig. 1).

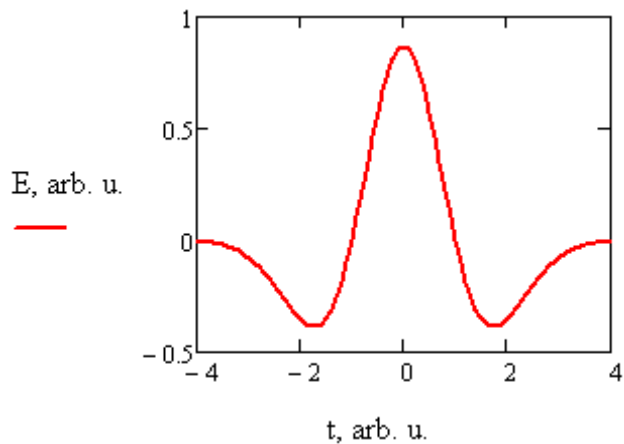

Fig. 1. A "Mexican hat" wavelet pulse

The spectrum of the pulse (1) is presented in Fig. 2. It is seen that the spectral width of the pulse is comparable to its center frequency, which is a characteristic feature of ultrashort pulses.

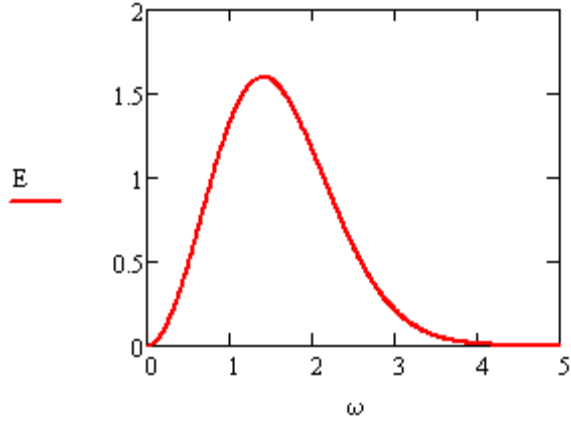

Fig. 2. The spectrum of the "Mexican hat" pulse

The position of the spectral maximum of this pulse is determined by the expression:

$$\omega_{max}^{(c)} = \frac{\sqrt{2}}{\Delta t}. \tag{2}$$

From this formula it also follows that $\omega_{max}^{(c)} \Delta t \approx 1$, that is, the pulse (1) can be approximately considered to be a half-cycle pulse if it is taken into account that the pulse duration is $\Delta t_p \simeq \sqrt{\pi}\,\Delta t$ for a pulse with a Gaussian envelope [5].

To calculate the propagation of an ultrashort electromagnetic pulse in a uniform isotropic medium with frequency dispersion in the linear approximation, we will use the following expression for the spatio-temporal dependence of the electric field strength in a propagating pulse:

$$E(t,z) = \int_{-\infty}^{\infty} E(\omega, z=0) \exp\left[i\omega(\tilde{n}(\omega)z/c - t)\right] d\omega/2\pi, \tag{3}$$

where $\tilde{n}(\omega)$ is the complex refractive index of the medium, $E(\omega, z=0)$ is the Fourier transform of the pulse at the initial point z=0. It is easy to check that the field (3) satisfies the wave equation in a medium with the specified dispersion law $\tilde{n}(\omega)$ and the boundary condition at the point z=0.

We will determine the complex refractive index of silver $\tilde{n}(\omega)$ within the framework of the Drude model by the formula for dielectric permittivity:

$$\varepsilon(\omega) = \varepsilon_\infty - \frac{\omega_p^2}{\omega(\omega + i\gamma)} = \varepsilon_\infty - \frac{\omega_p^2}{\omega^2 + \gamma^2} + i\frac{\gamma\,\omega_p^2}{\omega(\omega^2 + \gamma^2)}, \tag{4}$$

where $\gamma$ is the relaxation constant, $\omega_p = \sqrt{4\pi n e^2/m^*}$, $n$ is the conduction electron concentration, $m^*$ is the effective mass of an electron in the conduction band, $\varepsilon_\infty$ is the summand describing the contribution of bound electrons to the dielectric permittivity of a metal. In case of silver we have: $\hbar\omega_p = 9.1$ eV ($\omega_p = 1.38 \cdot 10^{16}$ $s^{-1}$), $\varepsilon_\infty = 3.7$, $\hbar\gamma = 18$ meV [6]. Then the function $\tilde{n}(\omega)$ is expressed in a standard manner in terms of the complex dielectric permittivity of the medium $\varepsilon(\omega)$.

The results of calculations by the above formulas are presented in Figs. 3-5. Plotted on the abscissa axis is the time in view of delay at the center frequency of the initial pulse $\tau = t - n'(\omega_{max})z/c$, where $n'(\omega)$ is the real part of the plasma refractive index, $c$ is the velocity of light in vacuum. The ordinate axis corresponds to the electric field strength in a pulse in relative units. In Figs. 3, 4 the parameter of duration of the wavelet pulse (1) is $\Delta t = 0.1$ $fs$. Then according to the equation (2) the center frequency of the cosine wavelet pulse (1) is $\omega_{max} = 1.5 \cdot 10^{16}$ $s^{-1}$ ($\lambda_{max} = 0.126$ $\mu m$), which exceeds the plasma frequency of silver $\omega_p$ by 9%.

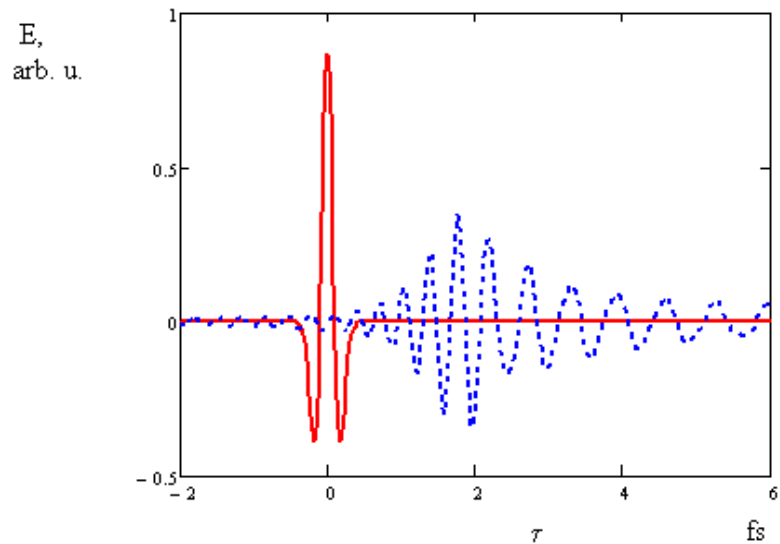

Fig. 3. The evolution of a cosine wavelet pulse: solid line - initial pulse, dotted line – pulse after propagation in silver to a distance of 1 μm

From Figs. 4-5 it follows that a given USP in silver spreads at distances of the order of several microns: as a result of dispersion spreading, a half-cycle pulse turns into a multicycle pulse. In this case the pulse amplitude decreases as well.

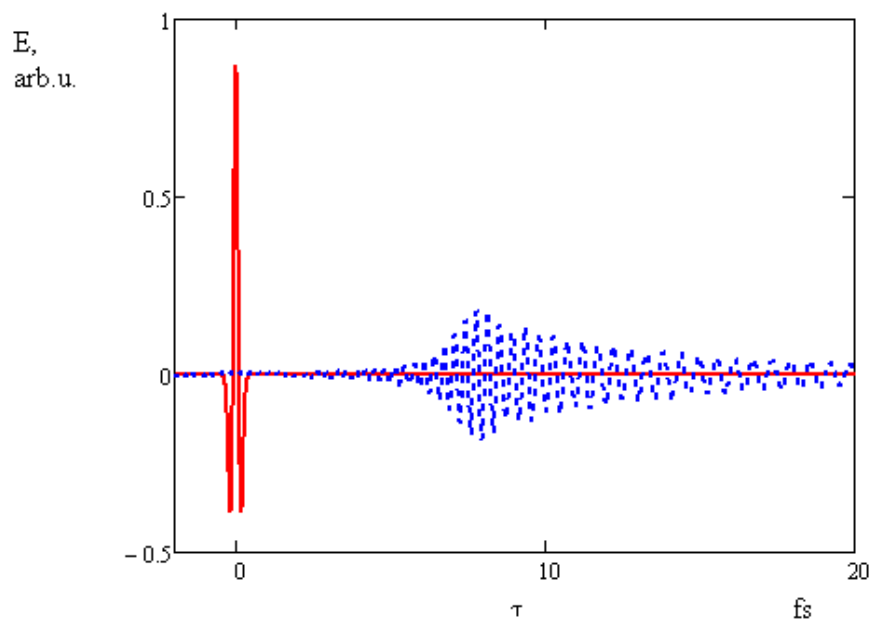

Fig. 4. The evolution of a cosine wavelet pulse: solid line - initial pulse, dotted line – pulse as a result of propagation in silver to a distance of 5 μm

Given in Fig. 5 is the comparison of dispersion spreading of the wavelet pulse (1) in silver for center frequencies that are higher and lower than the plasma frequency.

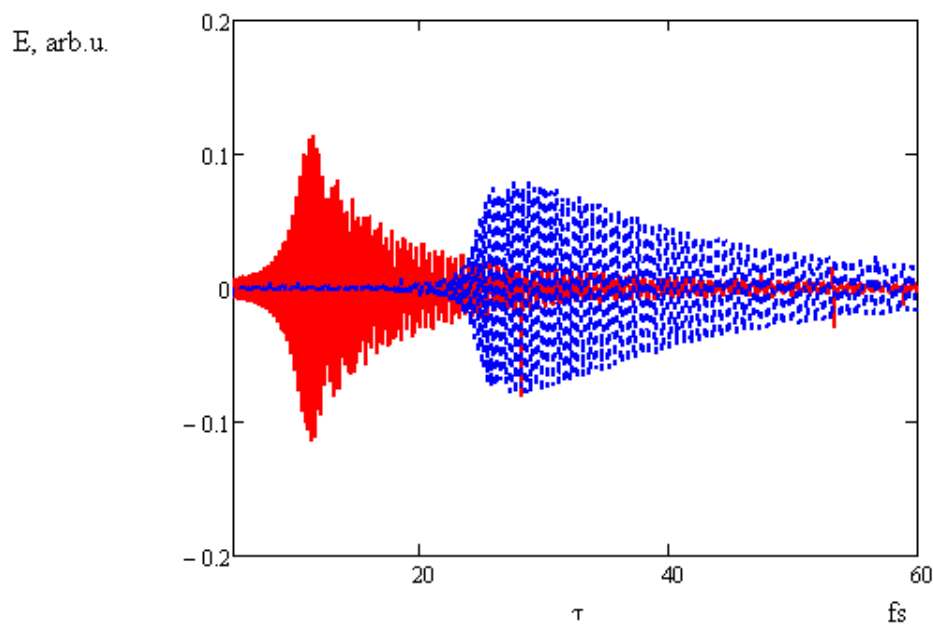

Fig. 5. The evolution of a cosine wavelet pulse with different center frequencies: solid line - $\omega_{max} = 1.8 \cdot 10^{16} \, s^{-1} > \omega_p$, dotted line – $\omega_{max} = 10^{16} \, s^{-1} < \omega_p$ as a result of propagation in silver to a distance of 10 μm

It is seen that for $\omega_{max} > \omega_p$ the dispersion delay caused by a large spectral width of the initial pulse is less than in the case $\omega_{max} < \omega_p$. Besides, spreading of a pulse with a lower center frequency is more appreciable, though the amplitudes of the pulses under consideration differ insignificantly.

The work has been done with the financial support of Russian Foundation for Basic Research (grant No. 13-02-00812).